\documentstyle[12pt]{article}

\newcommand\phibar{{\bar \phi}}
\newcommand\phat{\hat p}

\begin{document}

\begin{titlepage}
\noindent BUHEP-93-5 \hfill  July, 1993\\
\noindent hep-lat/9303014 \hfill Revised version\\

\begin{center}

{\LARGE\bf The Coleman-Weinberg Mechanism and First Order Phase Transitions}

\vspace{1.5cm}
\large{
Yue Shen\footnote{\noindent Email address: shen@budoe.bu.edu}\\
Physics Department, Boston University, Boston, MA 02215, USA\\
}
\vspace{1.9cm}
\end{center}

\abstract{
We study the Coleman-Weinberg phenomenon in a $U(N)\times U(N)$
symmetric scalar model in 4 dimensions.
By comparing the numerical simulation results with the bare perturbation
calculation in the weak bare coupling region, we
demonstrate explicitly that a first order transition is induced by
loop-effects. We also observed
first order phase transitions in the strong coupling region.
}
\vfill
\end{titlepage}

It is well known that a system with more than one relevant coupling
can have a first order phase transition induced by quantum loop fluctuations
-- a phenomenon first discovered by Coleman and
Weinberg \cite{Coleman,Halperin,Amit}.
The application of this Coleman-Weinberg mechanism led to a lower bound
of $7$-$10$ GeV on the Higgs mass in the Standard Model \cite{Weinberg,Ellis}
and to the suggestion that the QCD finite temperature chiral restoration phase
transition might be first order for more than two quark flavor\cite{Pisarski}.
It has also been suggested that the Coleman-Weinberg mechanism might be
relevant
for chiral symmetry breaking in technicolor models \cite{Paterson}.
Recently \cite{BU} it was pointed out that
the Coleman-Weinberg mechanism might be used to place constraints
on the top condensate model or strong-coupling extended technicolor models.

Let us consider the action (in Euclidean form)
\begin{equation}
S_\phi =
 \int d^4x \left\{ {1\over 2} tr \left(\partial_\mu \phi^\dagger\partial_\mu
\phi\right)
+ {m^2 \over 2}tr \left(\phi^\dagger\phi\right) +
\lambda_1\left(tr\phi^\dagger\phi\right)^2 + \lambda_2
tr\left(\phi^\dagger\phi\right)^2
\right\} ~,
\label{eq:scalar}
\end{equation}
for a scalar field $\phi$, where $\phi$ is a $N \times N$ complex matrix.
Such a model can be considered as a low energy effective action of
a top condensate model or strongly coupled extended technicolor model
(For simplicity, we ignore the fermions).
It has $U(N)\times U(N)$ global symmetry under transformation:
$\phi \to U\phi V^\dagger$,
where $U$ and $V$ are $N\times N$ unitary matrices. The dynamical scale
$\Lambda$ of a top condensate model or strongly coupled extended technicolor
model provides a natural cut-off. At energies much lower than $\Lambda$
the $U(N) \times U(N)$ symmetry is
spontaneously broken to $U(N)$. The vacuum expectation value (VEV)
$v$ associated with this symmetry breaking is supposed to be responsible for
generating the masses of the $W^\pm$ and $Z$.
For the model in Eq. (\ref{eq:scalar}) to be a good low energy effective
theory, there must be a large hierarchy between the VEV scale $v$ and the
cut-off scale $\Lambda$.

However, this model is expected to have
first order phase transitions because of the Coleman-Weinberg
mechanism \cite{Paterson,Rob}.
As $m^2$ is tuned through the critical point, the VEV jumps from zero in the
symmetric phase to a finite nonzero value in the
broken phase. In certain portions of parameter space the transition may
become strongly
first order such that the condition $v/ \Lambda \ll 1$ can no longer
be maintained. Therefore this part of the parameter space should be
excluded from the low energy effective model Eq. (\ref{eq:scalar}).
It is hoped that this in turn will constrain the parameter space of the
top condensate model or strong-coupling extended technicolor
model \cite{BU}.

For the convenience of later discussions we first reproduce
a geometric description of the
Coleman-Weinberg mechanism developed by Yamagishi \cite{Yama},
which combines the effective potential approach originally
used by Coleman and Weinberg and later developments in terms of
renormalization group (RG) flows and fixed point structure in parameter
space \cite{Amit,Paterson,Rob}.

In the parameter region $\lambda_2 >0$, spontaneous symmetry breaking
happens along the diagonal of matrix $\phi$ with a minimum at $\phi_{ij} =
\pm v \delta_{ij}$.
The tree level potential in a zero-mass theory can be written as
\begin{equation}
U(\varphi) = N(N\lambda_1 + \lambda_2) \varphi^4~,
\label{eq:tree_1}
\end{equation}
where $\varphi = \sqrt{tr(\phi^\dagger\phi)/N}$.
Away from $\varphi =0$ quantum fluctuation will contribute. However,
according to the solution of the RG equation for $U(\varphi)$,
the loop corrections can be included by letting $\lambda_1,\lambda_2$
run with $\varphi$ while keeping the form of $U(\varphi)$ in
Eq. (\ref{eq:tree_1}) unchanged (here we have accounted the fact that
the anomalous dimension, $\eta$, is zero in leading order for this model).
If we define $t = ln (\varphi/\mu)$ with $\mu$ a renormalization scale where
$\lambda_1, \lambda_2$ are defined, the running of $\lambda_1(t),\lambda_2(t)$
will be governed by $\beta$ functions
\begin{equation}
{d\lambda_1 \over dt} = \beta_1(\lambda_1, \lambda_2) ~,\ \ \ \ \
{d\lambda_2 \over dt} = \beta_2(\lambda_1, \lambda_2)
\end{equation}
with initial conditions $\lambda_1(0) = \lambda_1, \lambda_2(0) = \lambda_2$.
A local minimum will develop and the Coleman-Weinberg
phenomenon will be found if the running of $\lambda_1, \lambda_2 $ crosses
the ``stability line'' \cite{Yama}
\begin{equation}
4(N\lambda_1 + \lambda_2) + N\beta_1 + \beta_2 = 0 ~,
\label{eq:stab}
\end{equation}
in a region where
\begin{equation}
N\lambda_1 + \lambda_2 > 0~,\ \ \ \ \
4(N\beta_1+\beta_2) + N\sum_{i=1}^2 \beta_i {\partial \beta_1 \over \partial
\lambda_i}
+ \sum_{i=1}^2 \beta_i {\partial \beta_2 \over \partial
\lambda_i} >0 ~.
\label{eq:U0_1}
\end{equation}

In the region $\lambda_2 <0$ the symmetry breaking
happens along one diagonal element $\phi_{ij} = \pm v \delta_{1i}
\delta_{1j}$. In this case we write the tree level potential as
\begin{equation}
U(\varphi) = (\lambda_1 + \lambda_2) \varphi^4~,
\label{eq:tree_2}
\end{equation}
with $\varphi = \sqrt{tr(\phi^\dagger\phi)}$.
Again the loop corrections can be included in the running of $\lambda_1,
\lambda_2$. The Coleman-Weinberg phenomenon will occur if the running of
$\lambda_1, \lambda_2$ crosses the ``stability line''
\begin{equation}
4(\lambda_1 + \lambda_2) + \beta_1 + \beta_2 = 0 ~,
\label{eq:stab2}
\end{equation}
in a region where
\begin{equation}
\lambda_1 + \lambda_2 > 0~, \ \ \ \ \
4(\beta_1+\beta_2) + \sum_{i,j=1}^2 \beta_i {\partial \beta_j \over \partial
\lambda_i}
 >0 ~.
\label{eq:U0_2}
\end{equation}

Using the one-loop perturbative $\beta$-functions \cite{Paterson,Rob}
\begin{eqnarray}
\pi^2{d\lambda_1 \over dt} &=& (N^2 + 4)\lambda_1^2 + 4N\lambda_1\lambda_2 +
3\lambda_2^2~, \\ \nonumber
\pi^2{d\lambda_2 \over dt} &=& 6\lambda_1\lambda_2 + 2N\lambda_2^2 ~,
\label{eq:beta}
\end{eqnarray}
the RG flow of $\lambda_1, \lambda_2$ is plotted in Figure 1 for $N=2$.
Notice that
the arrows in Figure 1 indicate the direction of {\it decreasing} $\varphi$
(i.e. the infrared limit of the theory). The ``stability lines'' are indicated
by the dotted lines. They are entirely inside the regions that satisfy
the conditions of Eqs. (\ref{eq:U0_1}) and
Eqs. (\ref{eq:U0_2}). Thus whenever the flow lines cross the
``stability line'' one expects the Coleman-Weinberg
phenomenon\footnote{\noindent In Figure 1,
the perturbative ``stability lines'' terminate in the strong
coupling region (solid circles).
This may merely mean the breaking down of perturbation theory.
Indeed, we find in numerical simulation a
first order phase transition at $\lambda_1 = -19.0, \lambda_2=40.0$.}.
In the language of the RG flow the Coleman-Weinberg phenomenon occurs due to
the
absence of an infrared stable fixed point. The system is driven to a runaway
RG trajectory in the infrared limit and the vacuum at $\varphi = 0$ becomes
unstable \cite{Amit,Paterson,Rob}.

The  above description is simple and elegant. However,
the $\beta$-functions and consequently the RG flow in Figure 1 was
calculated in the leading
order renormalized perturbation theory.
In principle it is possible that there might
exist a nonperturbative infrared stable fixed point and
the phase transition could become second order inside its attractive domain.
For the purpose of
Ref \cite{BU}, this is an important question. If one stays around
$\lambda_1 = \lambda_2 = 0$ the couplings run very slowly. It takes many
decades to reach the VEV scale from the cut-off scale. Then the phase
transition becomes practically second order and there can be a large
hierarchy between $\Lambda$ and the VEV.
It is only in the region where $\lambda_1$
is small and $\lambda_2$ is relatively large that the RG flow becomes fast
and a small $\Lambda / v$ ratio becomes likely.
This has been used \cite{BU} as an argument to
exclude the large $\lambda_2$
region from the model in Eq. (\ref{eq:scalar}).
However, should there be an infrared stable fixed point in the large parameter
region, the model in Eq. (\ref{eq:scalar}) would have practically
no constraints on its parameter values (besides the obvious
condition $N\lambda_1 + \lambda_2 > 0$ for the bare action to be stable).
Therefore, it was suggested \cite{BU} that a numerical simulation
is needed to look for first order phase transitions in the strong
coupling region.

Interest in numerical studies of the Coleman-Weinberg phenomenon
has also been
raised in a recent work of March-Russell \cite{March}. He considered
gauge-Higgs systems in 3 dimensions which are thought to have
first order phase transitions due to the Coleman-Weinberg mechanism.
He argued that the conventional picture based on the $d=4-\epsilon$
expansion result is unreliable for $\epsilon=1$
and the phase transition for such systems
in 3 dimensions could be
second order\footnote{\noindent Numerical simulations of a spin model
\cite{Bak}
and $SU(N)$, $U(N)$ nonlinear sigma models \cite{Kogut} in 3 dimensions
found evidences of first order phase transitions. These works seem to favor
the $\epsilon$-expansion predictions. }.
The model in Eq.~(\ref{eq:scalar})
can serve as a testing ground for this conjecture
\footnote{\noindent A Monte-Carlo simulation had been performed before
for the model in Eq.~(\ref{eq:scalar}) in 3 dimensions \cite{Gaust}.
Unfortunately
the result of Ref. \cite{Gaust} was wrong. The point shown to have two
coexisting phases in that paper (Figure 2) actually is deep in the broken
phase. Our investigation of this model in 3 dimensions found
first order phase transitions and the results will be reported
elsewhere.}.

In 4 dimensions there is no problem associated with the $\epsilon$-expansion
and the Coleman-Weinberg phenomenon is expected to exist at least in the
weak coupling region. However, clear numerical evidence for such phenomenon
has not been obtained so far.
In the following, we report our first results in the numerical
investigation of the model in Eq.~(\ref{eq:scalar}) in 4 dimensions.

For the numerical simulations we set $N=2$ in Eq. (\ref{eq:scalar}).
The simple Metropolis
algorithm is used with a uniform step size $\Delta \phi_{ij} = \Delta \phi$
which is tuned such that the acceptance rate is around $60\%$.
Since there can be no symmetry breaking in a finite volume it is important
to choose an order parameter invariant under $U(N)\times U(N)$.
For a lattice with linear size $L$, we define
\begin{equation}
\phibar_{ij} = {1\over L^4}\sum_x \phi_{ij}(x)~,
\end{equation}
and use $\langle tr(\phibar^\dagger\phibar)\rangle$
as the order parameter.
This order parameter is a finite number in the broken
phase and vanishes like $O(1/L^4)$ in the symmetric phase.

For this exploratory investigation most simulations were done on relatively
small lattices ($4^4, 6^4$ and $10^4$).
To decide the order of phase transitions we looked for the hysteresis effects
in the thermocycles: we perform a series of runs back and forth across the
critical region.
Each run uses the last configuration of the previous run as its initial
configuration. In case of a first order phase transition, supercooling
or superheating will produce the hysteresis effects.
In addition, we looked at
the histogram for the distribution of $tr(\phibar^\dagger\phibar)$.
Since the histogram distribution is proportional to $\exp \{-L^4
U_L(\phibar)\}$
, where $U_L(\phibar)$ is the effective potential,
one should observe a double peak structure throughout the critical
region in a first order phase transition. There are more elaborate
techniques \cite{Bock} to determine the order of phase transitions
which will not be used here.

We choose the parameters $\lambda_1$ and $\lambda_2$ such that
the bare lattice action is absolutely stable
\begin{equation}
\lambda_2 > 0~, \ \ \ N\lambda_1 + \lambda_2 >0 ~.
\end{equation}
According to the RG flow in Figure 1, $\lambda_1$ and $\lambda_2$ run
very slowly in the
weak coupling region. Qualitatively, a weak coupling bare action fixed to the
right of the ``stability line" in Figure 1, will need many decades of running
in energy scale to get a renormalized theory to cross the ``stability line".
The running becomes fast in the strong $\lambda_2$ region.
On the lattice the running is controlled by the correlation length $\xi$
with $t \approx \ln \xi$, which is limited by the
lattice size $\xi
\raise.3ex\hbox{$<$\kern-.75em\lower1ex\hbox{$\sim$}} L$.
Therefore, to observe the Coleman-Weinberg phenomenon
one either has to perform simulations on a large enough
lattice \footnote{We believe this might be the physical reason for the
phenomenon observed in Ref. \cite{Kogut}. }
or has to move the bare theory very close to the ``stability line'' in
order to get a renormalized theory to cross the ``stability line''.

Our numerical results in the weak coupling region are shown in Figures 2 and 3
where the bare parameters are chosen to be
very close to the ``stability line'' Eq. (\ref{eq:stab}).
The phase transition is clearly
first order at $\lambda_1=-0.22, \lambda_2 = 0.5$.
Figure 2 shows thermocycles for the $4^4, 6^4$ and $10^4$ lattices.
For the $4^4$ lattice each point in the thermocycle has 20000 sweeps
as warm-up and 60000 sweeps in the measurement. For the $6^4$ and $10^4$
lattices they are 3000, 10000 and 400, 1200, respectively. Away from the
critical point the finite size effects are small.
They become larger close to the critical point:
the hysteresis effect is barely
visible on the $4^4$ lattice and becomes stronger on larger lattices as shown
in Figure 2.
Figure 3 shows the tunnelings between the broken and symmetric phases on the
$4^4$ lattice at the ``critical point''.

In the weak coupling region one may calculate the order parameter
$\langle tr(\phibar^\dagger\phibar)\rangle$ in the bare perturbation expansion
and compare with the numerical results.
In particular,
for $\lambda_2 >0$
the effective potential $U_L(\phibar)$ has a global minimum
at $\phibar_{ij} = \pm v \delta_{ij}$. If we define
$\varphi \equiv \sqrt{tr(\phibar^\dagger\phibar)/N}$,
the one-loop effective potential is given by
\begin{eqnarray}
U(\varphi) &=& N\left({m^2 \over 2}\varphi^2
+ (N\lambda_1+\lambda_2)\varphi^4\right)
+{1\over 2L^d}\sum_{p\ne 0} \ln\left[ \phat^2+m^2+12(N\lambda_1
+\lambda_2)\varphi^2\right]\\ \nonumber
&+&{N^2-1\over 2L^d}\sum_{p\ne 0}
\ln\left[ \phat^2+m^2+4(N\lambda_1+3\lambda_2)\varphi^2\right]
+{N^2\over 2L^d}\sum_{p\ne 0} \ln\left[ \phat^2+m^2+4(N\lambda_1+\lambda_2)
\varphi^2\right]~,
\label{eq:efpt}
\end{eqnarray}
where $d =4$ and $\phat^2 = \sum_\mu \sin^2p_\mu$.
In the infinite volume limit the lattice sum should be replaced by a
momentum space integral
${1 \over L^d}\sum_{p \ne 0} \to \int {d^d p \over (2\pi)^d}$.

At tree-level the effective potential Eq.~(12) 
appears to indicate a second order
phase transition at $m^2 = 0$ for arbitrary values of $\lambda_1,\lambda_2$
(provided $N\lambda_1 + \lambda_2 >0$).
This is, however, changed at one-loop.
For example, if we plot the one-loop effective potential
at $\lambda_1 = -0.22, \lambda_2 = 0.5$ for various $m^2$ values.
In contrast to the tree-level
picture where there is always a single minimum, we will have two minima at
$\varphi=0$ and
$\varphi =v > 0$. As $m^2$ is tuned from $-1.0$ to $-1.1$ the second minimum at
$\varphi>0$ changes from a local minimum to a global minimum: a picture
ordinarily found at a first order phase transition.

The dotted line in Figure 2
indicates $\langle tr(\phibar^\dagger\phibar)\rangle$ as calculated in
the bare perturbation theory.
The agreement with the numerical result is reasonably well considering
that $\lambda_2=0.5$ is not exactly weak coupling with our normalization.
Numerically we found that the bare perturbation theory
works within $20\%$ up to $\lambda_1, \lambda_2 \approx 1$.

Since very often first order phase transition happens at a finite
correlation length where the cut-off effects can not be neglected (known as
``no continuum limit''),
one may wonder why the description of the Coleman-Weinberg phenomenon
in the renormalized language could be qualitatively correct.
The answer lies in the fact that the Coleman-Weinberg phenomenon happens
in the infrared. Given a renormalized theory $\lambda_{1R}, \lambda_{2R}$
fixed to the left of the ``stability line" in Figure 1, one can always
arrange (at least in the small coupling region) the bare theory sufficiently
far away from the ``stability line"
along a trajectory such that it takes many decades in energy scale for
the bare theory to ``run" into the renormalized theory. In such arrangement
the cut-off effects can be made small (the phase transition becomes weakly
first order) and the use of the renormalized theory
language becomes valid.

Of course, the cut-off effects can not be neglected
in a parameter region where the phase transition is strongly first order.
As pointed out in Ref. \cite{BU}, such region must be excluded from the
continuum theory because the renormalized theory is no longer
meaningful\footnote{This is similar in spirit to the argument for the
triviality bound on the Higgs mass \cite{Peter}. To actually map out the
excluded region in the $(\lambda_1,\lambda_2)$ parameter space requires
detailed work and is under investigation.}.

In the strong coupling region we also found first order
phase transitions. For example,
in Figure 4 we show our simulation results at $\lambda_1=0, \lambda_2=10$
on the $10^4$ lattice.
The lower and upper curve correspond to Monte Carlo runs with disordered and
ordered initial conditions.
The fact that we are able to observe a first order phase transition
here on a relatively small lattice conforms to the qualitative
feature of the RG:
although the point $\lambda_1=0, \lambda_2=10$ appears to be far away
from the ``stability line" in Figure 1, the running is fast
due to the strong $\lambda_2$ coupling.

In conclusion, by comparing the numerical simulation results with the
one-loop bare perturbation calculation in the weak coupling region,
we have shown explicitly
that in $d=4$ quantum fluctuations can induce a
first order phase transition--commonly known as the Coleman-Weinberg phenomenon
\cite{Coleman}.
We also found first order phase transitions in the strong $\lambda_2$
coupling region. This might be taken as an indication that there is no
infrared stable fixed point in the nonperturbative region. However,
a systematic search (which is under investigation) is needed in order
to confirm this point.

The numerical work to determine the complete phase diagram
in both weak and strong coupling regions will be
reported elsewhere \cite{future}.

\noindent{\it Acknowledgements}

I thank S. Chivukula for useful discussions. I also thank him and
E. Simmons for carefully reading and correcting my manuscript.
This work was supported in part under DOE contract DE-FG02-91ER40676 and NSF
contract PHY-9057173, and by funds from the Texas National Research Laboratory
Commission under grant RGFY92B6.

\pagebreak

\pagebreak

\section*{Figure Captions}

\noindent{\bf Figure 1}: Solid lines indicate the RG flow
in $\lambda_1,\lambda_2$
parameter space. The arrows point to the infrared direction. The dotted lines
are the ``stability lines'' calculated according to Eqs. (\ref{eq:stab})
and (\ref{eq:stab2}) using the one-loop $\beta$-functions.
They stop at end points (solid circle).

\noindent{\bf Figure 2}: $\langle tr(\phibar^\dagger\phibar)\rangle $
as a function of $m^2$
at $\lambda_1 = -0.22, \lambda_2 = 0.5$. The open circles indicate
data on the $4^4$ lattice and open squares and open triangles are for
$6^4$ and $10^4$ lattice respectively.
Unless explicitly shown the statistical errors for data points are smaller
than the size of symbols. The solid lines connect data points
from $6^4$ and $10^4$ lattices to indicate the thermocycles.
The dotted line gives the bare perturbation prediction.

\noindent{\bf Figure 3}: Time history shows tunneling between two states
at $\lambda_1 = -0.22, \lambda_2 = 0.5, m^2 = -0.875$. Data is obtained
on the $4^4$ lattice. Tunneling becomes rare on larger lattices.

\noindent{\bf Figure 4}: Time history shows two-states signal on the $10^4$
lattice at $\lambda_1 = 0, \lambda_2 = 10, m^2 = -29.625$. The lower and
upper curve correspond to runs with disordered and ordered initial conditions.

\end{document}